\documentclass[reprint, 
amsmath,amssymb,aps,prl,superscriptaddress
]{revtex4-2}

\usepackage{graphicx}
\usepackage{dcolumn}
\usepackage{bm}
\usepackage{comment}
\usepackage{xcolor}
\usepackage[normalem]{ulem}
\usepackage{braket}

\begin{document}

\preprint{APS/123-QED}

\title{Pushing Photons with Electrons: Observation of the Polariton Drag Effect}

\author{D. M. Myers}
\thanks{Current address: Kulicke \& Soffa Industries, Inc., Fort Washington, PA 19034, USA}
\affiliation{Department of Physics and Astronomy, University of Pittsburgh, Pittsburgh, PA 15260, USA}

\author{Q. Yao}
\thanks{Current address: Joint Quantum Institute, University of Maryland, College Park, Maryland 20742, USA}
\email{qiyao@umd.edu}
\affiliation{Department of Physics and Astronomy, University of Pittsburgh, Pittsburgh, PA 15260, USA}

\author{H. Alnatah}
\affiliation{Department of Physics and Astronomy, University of Pittsburgh, Pittsburgh, PA 15260, USA}

\author{S. Mukherjee}
\thanks{Current address: Joint Quantum Institute, University of Maryland, College Park, Maryland 20742, USA}
\affiliation{Department of Physics and Astronomy, University of Pittsburgh, Pittsburgh, PA 15260, USA}

\author{B. Ozden}
\thanks{Current address: Department of Physics and Engineering, Penn State Abington, Abington, Pennsylvania 19001, USA}
\affiliation{Department of Physics and Astronomy, University of Pittsburgh, Pittsburgh, PA 15260, USA}

\author{J. Beaumariage}
\affiliation{Department of Physics and Astronomy, University of Pittsburgh, Pittsburgh, PA 15260, USA}
\author{L. N. Pfeiffer}
\affiliation{Department of Electrical Engineering, Princeton University, Princeton, NJ 08544, USA}

\author{K. West}
\affiliation{Department of Electrical Engineering, Princeton University, Princeton, NJ 08544, USA}

\author{D. W. Snoke}
\affiliation{Department of Physics and Astronomy, University of Pittsburgh, Pittsburgh, PA 15260, USA}

\date{\today}

\begin{abstract}
We show the direct effect of free electrons colliding with polaritons, changing their momentum. The result of this interaction of the electrons with the polaritons is a change in the angle of emission of the photons from our cavity structure. Because the experiment is a photon-in, photon-out system, this is equivalent to optical beam steering of photons using a direct electrical current. The effect is asymmetric, significantly slowing down the polaritons when they move oppositely to the electrons, while the polariton momentum only slightly increases when electrons moving in the same direction. We present a theoretical model which describes this effect. 
\end{abstract}

\maketitle

It has long been known that an electron can impart momentum to an exciton-polariton through the scattering interaction with the exciton part of the polariton~\cite{parish}.  Apart from the intrinsic interest in demonstrating this effect, it may have application in steering the direction of light emission by a direct electrical current, with the angle of deflection directly controlled by the applied current. However, observation of this effect has remained elusive. In this Letter we report observation of this effect with exciton-polaritons in a solid-state microcavity.  

The physics of exciton-polaritons has been widely explored in recent years, and is well summarized in recent reviews, many of which have focused on Bose-Einstein condensation of polaritons \cite{kasprzak2006bose,abbarchi2013macroscopic,sanvitto2010persistent,lagoudakis2009observation,kavokin-book,CarusottoRev2013,Deng2010,SnokeKeelingPT2017}. In particular, there has been great interest in one-dimensional wires for conducting polariton transport \cite{bloch1,bloch2,kavring}. In the present work, we use a polariton condensate, but the effect is not fundamentally one that only occurs for a condensate; rather, the condensate produces a spectrally narrow emission over a wide area that makes the polariton drag effect easy to observe. Superfluidity of the condensate does not prevent the drag effect, because it is equivalent to a body force, or longitudinal force. The structures we use also allow long-distance transport over hundreds of microns \cite{Nelsen2013, Steger2015}, which allows good momentum resolution in our measurements. 
 
 Recent theoretical work taking into account the polaron effect on an exciton interacting with a Fermi sea \cite{CotletPolaronDrag2018} has predicted that applied direct current will give a drag force on exciton-polarons, and that the effect will also be significant when the excitons are coupled to photons, as in an exciton-polariton system \cite{Sidler2017}. However, the drag effect can be understood even apart from the polaron effect as a purely collisional exchange of momentum due to Coulombic collisions between electrons and the excitonic part of a polariton, as discussed, e.g., in Ref. \onlinecite{hartwell} and references therein. Other work \cite{BermanPRB2010, BermanPhysLetA2010, BermanACSNano2014} envisioned a drag effect between separated layers of excitons and free electrons, but this is not crucial for the effect, as free electrons can also move in the same quantum wells as the excitons.

{\bf Experimental Method}. Exciton-polaritons in the strong-coupling limit were made by placing semiconductor quantum wells at the antinodes of a planar optical cavity. These semiconductor layers have excitons nearly resonant in energy with a cavity photon mode. If the $Q$-factor of the cavity is high enough, and the coupling of the photon and exciton states is strong enough, new eigenstates appear which are no longer purely photon or exciton, but a superposition of both. In other words, a photon in the cavity spends some fraction of its time as an excited electron-hole pair. 

The sample used in this experiment consists of a $3\lambda/2$ microcavity formed by two distributed Bragg reflectors (DBRs), grown by molecular beam epitaxy (MBE). The DBRs were both made of alternating layers of Al$_\mathrm{0.2}$Ga$_\mathrm{0.8}$As and AlAs, with 32 periods in the top DBR and 40 in the bottom. 4 quantum wells (QWs), made of 7 nm GaAs layers with AlAs barriers, were placed at each of the 3 antinodes of the cavity. This microcavity design is the same used in previous work \cite{Nelsen2013, Steger2015, Myers2017, Ozden2018, MyersPRB2018, Jonny2024} (see the Supplemental Material~\cite{SI} for more details of the sample characteristics). Long wires ($\approx 200~\mu\mathrm{m} \times 20~\mu\mathrm{m}$) were formed by etching away the top DBR, confining the polaritons within the wire and exposing the QWs. 

\begin{figure}
\centering
\includegraphics[width=\linewidth]{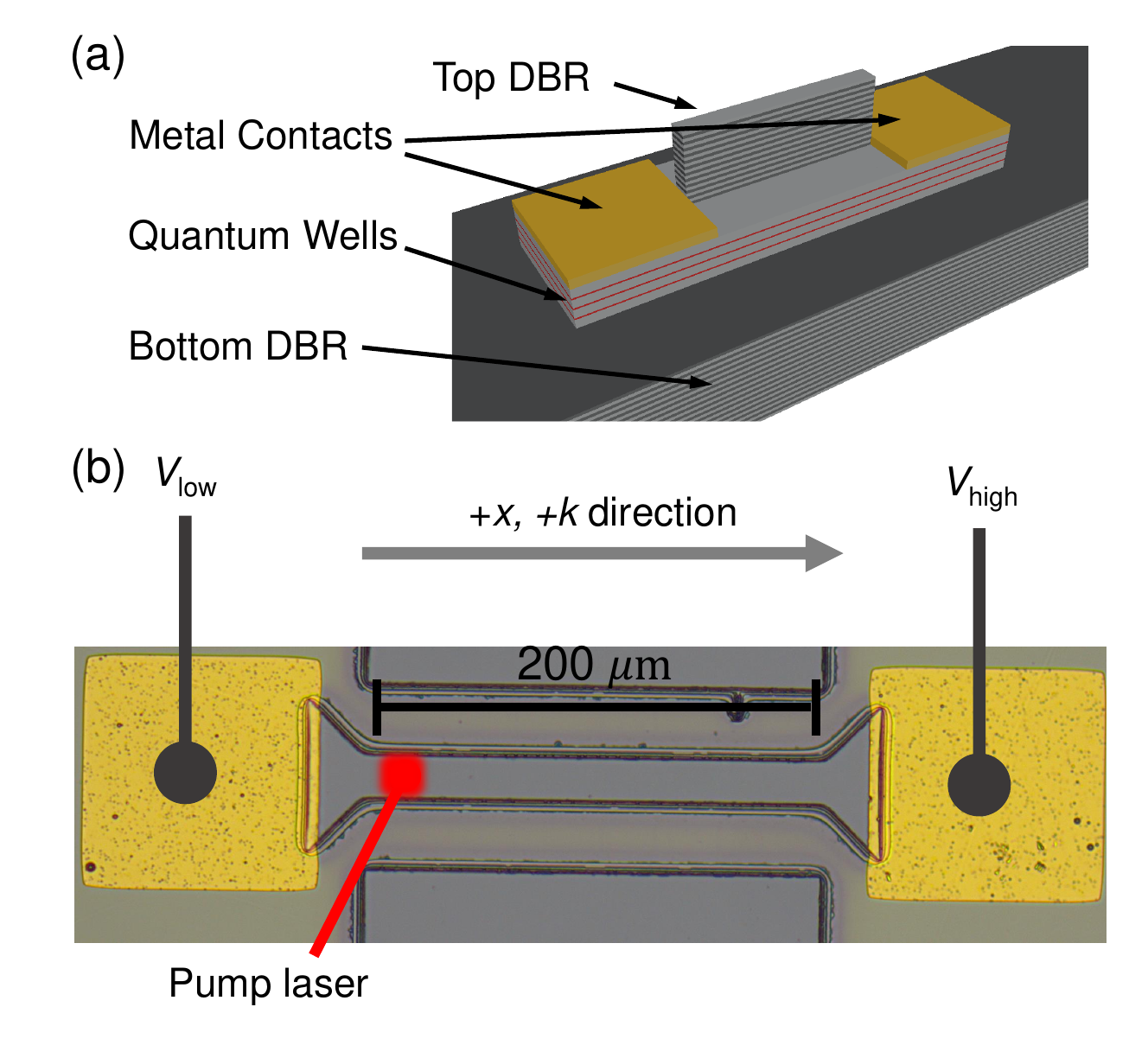}
\caption{(a) The etched microcavity structure with the NiAuGe contacts at each end of the wire. (b) Optical image of a representative etched wire, with an overlay showing the experimental arrangement. The straight wire part is 200 $\mu$m long. The voltage connections of the source were connected to the device as shown. Thus, positive conventional current flowed in the direction of $-x$.}
\label{fig:Diagram}
\end{figure}

To allow co-linear motion of free electrons in the same medium, two NiAuGe contacts were then placed upon the QWs at the ends of the wire, allowing electrical injection into the QWs (see Figure \ref{fig:Diagram}). Hall measurements show that the wire regions are intrinsically $n$-doped of the order of $10^{13}$~cm$^{-3}$, and the contacts are $n$-type with heavy doping of the order of $10^{19}$~cm$^{-3}$. A continuous-wave (CW) stabilized M Squared Ti:sapphire pump laser was used to excite a spot on the straight part of the wire, with a spot size with diameter about 20~$\mu$m, in the same arrangement used in Ref. \onlinecite{MyersPRB2018}. The pump was non-resonant, with an excess energy of about 100 meV, and mechanically chopped at 400 Hz with a pulse width of about $60~\mu$s. A source meter was used to sweep the applied voltage along the wires while measuring electrical current. The details of the fabrication and the electrical measurement are discussed in Ref. \onlinecite{Ozden2018}. The cavity detuning of the wire device presented in the main paper was $\delta = 6.8\; \mathrm{meV}$, corresponding to an exciton fraction $|X|^2 = 0.67$ for the lower polariton at $k = 0$. A full characterization of the sample is presented  in the Supplementary Material~\cite{SI}.

Off-resonant optical pumping of this type of structure produces a cloud of excitons, which then lose energy and fall down into polariton states. As in previous experiments with similar structures \cite{Myers2017, MyersPRB2018}, there are two density thresholds for the optical excitation. At the lower threshold, a local quasicondensate is formed at the excitation spot, which can then ballistically expand away from the excitation region; above a higher critical threshold, the condensate jumps down dramatically into a much lower energy state, which is spectrally very narrow ($< 0.1$~meV width at half maximum), has high coherence length ($\gg 200$~$\mu$m), and fills most of the available space in a potential-energy minimum. We define the pump power needed to reach this second threshold as $P_{\rm thres}$. The condensate was observed by recording the photons that leak out of the top mirror, using conventional imaging optics for both real-space (near field) and momentum-space (far field, Fourier plane) images, and a spectrometer for energy resolution. The leakage of photons out of the cavity was a tiny fraction of the population at any moment in time, because the $Q$ of the cavity is very high ($\sim 350,000$), so that a steady-state population was maintained \cite{Steger2015}. The long polariton lifetime of 300-400 ps allows macroscopic flow over hundreds of microns \cite{Steger2015} and demonstration of true equilibrium, confirmed by fits of the particle occupation number as a function of energy to an equilibrium Bose-Einstein distribution \cite{sun2017bose,alnatah2024coherence}. In this steady state, the energy and spatial distribution of the polariton condensate is determined by the external potential profile it feels, which is a combination of the single-particle polariton dispersion and the repulsion of polaritons from slow-moving excitons with much higher mass, and the density-dependent, repulsive polariton-polariton interaction, which tends to flatten any external potential felt by the polaritons.

\begin{figure}
\centering
\includegraphics[width=\linewidth]{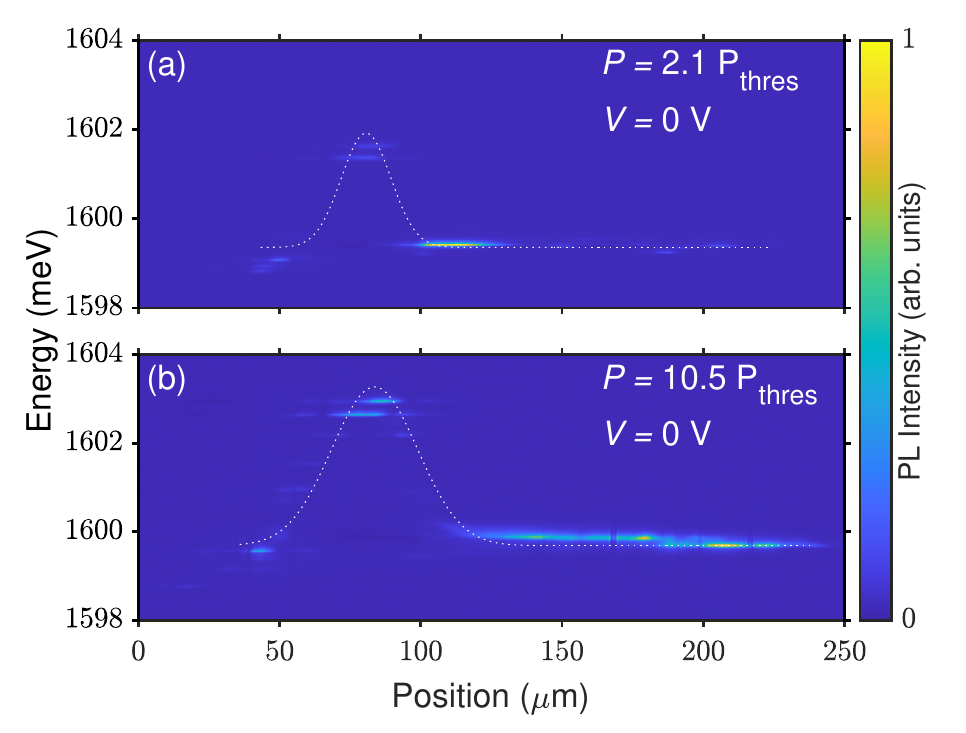}
\caption{Photoluminescence intensity vs. energy and position along the length of the wire at zero applied voltage. The white dotted lines are reference lines, which give the outline of the potential felt by the polaritons. The threshold power ($P_\mathrm{thres}$) was about 75 mW, and the powers used were (a) 2.1$P_\mathrm{thres}$ and (b) 10.5$P_\mathrm{thres}$. The PL intensity was normalized separately for each image.}
\label{fig:EvR}
\end{figure}

{\bf Experimental Results}. The experimental assignment of the direction of positive $x$ and $k$ relative to the electrical contacts and pump spot is shown in Figure 1(b). Figure \ref{fig:EvR} shows the polariton distribution along the wire at two different pump powers above threshold and at zero applied voltage. The energy ``hill'' near 75~$\mu$m is due to the large population of excitons at the pump spot, which repulsively interact with the polaritons. 
A shallow cavity gradient along the $+x$-direction caused a small slope in the potential, with a total energy drop of about 0.5 meV across the whole wire. At the lower power, 
a large mono-energetic condensate forms on the side away from the pump, filling a region of the wire about 100~$\mu$m long. At higher power, the population on the right becomes large enough to have repulsive interactions that shift it up in energy, forming a single, nearly mono-energetic condensate.

\begin{figure}
\centering
\includegraphics[width=\linewidth]{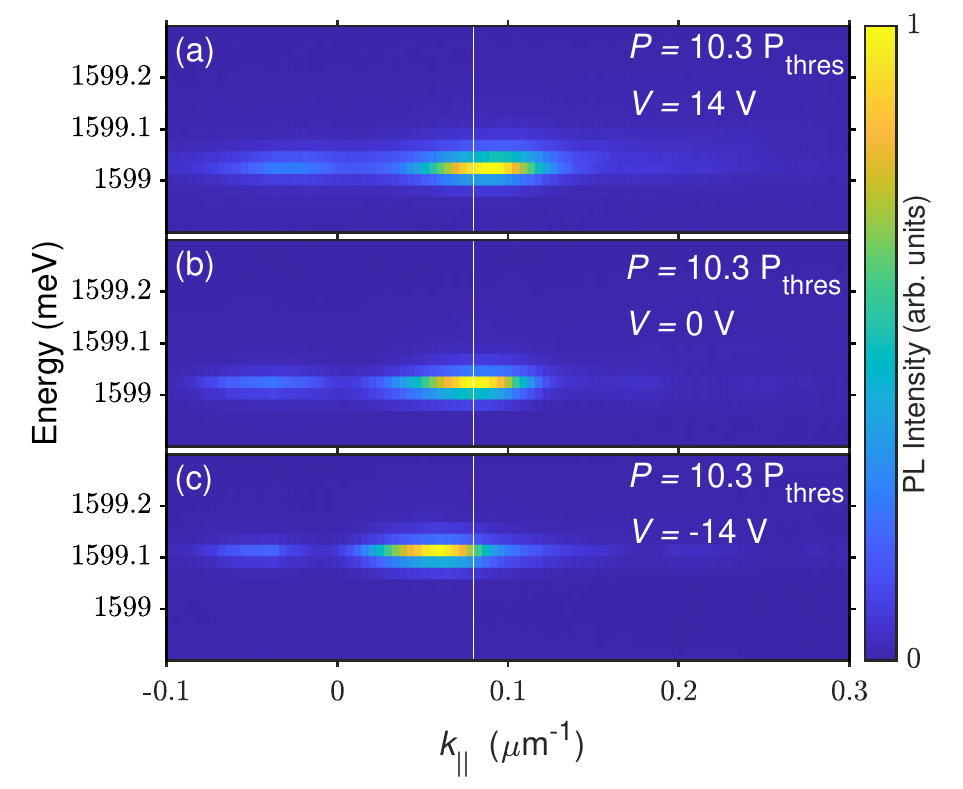}
\includegraphics[width=0.9\linewidth]{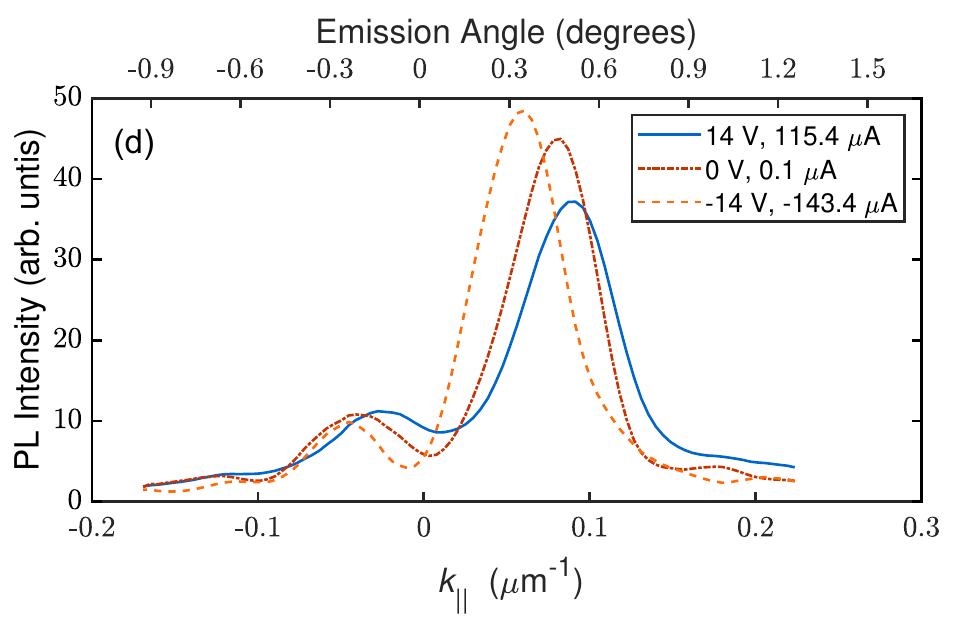}
\caption{Photoluminescence intensity vs. energy and in-plane momentum ($k_\parallel$) for the polariton condensate in the wire, excluding the pump region, under three different applied voltages. The threshold power ($P_\mathrm{thres}$) was about 75 mW, and the powers used were about 10.3$P_\mathrm{thres}$. The applied voltage in each case (a) 14~V and (b) 0~V (c) -14~V. The white vertical line marks the $k_\parallel$ value of the peak of the distribution at zero applied voltage to aid in comparing the images. The PL intensity was normalized separately for each image. (d) Time integrated average momentum distribution of the condensate under different applied voltages derived from (a), (b), and (c).}
\label{fig:EvK}
\end{figure}

Figure \ref{fig:EvK} shows the energy vs. in-plane momentum ($k_\parallel$) of the polaritons in the level region, which is on the right side of pump spot (Figure \ref{fig:EvR}), at three different applied voltages. 
With zero applied voltage (Figure \ref{fig:EvK}(b)), an overall nonzero in-plane momentum is observed, which is due to flow in the main part of the wire, away from the pump spot, in the $+x$-direction. With negative applied voltage (Figure \ref{fig:EvK}(c)), while all the other experimental conditions remain the same, the overall momentum is clearly reduced. This voltage corresponds to conventional current flowing in the $+x$-direction, and thus electron current flowing in the $-x$-direction, which opposes the polariton flow. With positive applied voltage (Figure \ref{fig:EvK}(a)), the overall momentum is increased. This clearly shows an effect of drag upon the polaritons from the electrical current.

Figure \ref{fig:EvK}(d) shows the polariton distribution vs.~$k_\parallel$ for multiple applied voltages. A clear shift in the momentum is observed under applied voltages. Simple calculations indicate that the electrons move slower than the polaritons. The polariton steady-state velocity is measured directly from their momentum; 0.087~$\mu$m$^{-1}$ corresponds to $v = \hbar k/m = 5.6\times10^6$~cm/s. The electron velocity can be estimated from the mobility; Hall measurements on these samples at room temperature give mobility of approximately 100~cm$^2$/V-s. For a voltage drop of 15~V over 200~$\mu$m, this gives electron velocities of the order of $7.5\times10^4$~cm/s, which is expected to increase at low temperature \cite{pfeifferref}, as implied by the fits to the data discussed below.

\textbf{Theory}. We have developed a quantum Boltzmann model showing the drag effect originating from the collisions between polartions and electrons. We consider polariton-electron scattering where the two particles scatter from the state $\ket{\vec{k}_i} = \ket{k_1\vec{q}_1}$ to state $\ket{\vec{k}_f} = \ket{k_2\vec{q}_2}$ through a scattering interaction $V_{\text{int}}$. The interaction potential is given by
\begin{equation}\label{eq:Vint}
V_{\text{int}} = V_{0}\sum_{k_1,k_2,\vec{q}_1}a^{\dagger}_{k_2}b^{\dagger}_{\vec{q}_2}b_{\vec{q}_1}a_{k_1},    
\end{equation}
where $a^{\dagger}$ and $a$ are the polariton creation and annihilation operators respectively and $b^{\dagger}$ and $b$ are the electron creation and annihilation; $\vec{q}_2$ is determined by momentum conservation. The drag force that the electrons exert on a polariton with momentum $k_1$ is calculated using Fermi's Golden rule
\begin{eqnarray}\label{eq:F_k}
F(k_1) &=&  2 \pi V^2_{0} \sum_{k_2, \vec{q}_2}  \left ( k_2-k_1 \right )N_e(\left | \vec{q}_1-\vec{q}_0 \right |)\\
&&\times  \left ( 1+N_p(k_2-k_0) \right)\delta\left(E_{k_2}+E^e_{\vec{q}_2}-E_{k_1}-E^e_{\vec{q}_1}\right). \nonumber
\end{eqnarray}
The wavevectors $k_1$ and $k_2$ correspond to the incoming and outgoing polariton with mass $m_p$, respectively and $\vec{q}_1$ and $\vec{q}_2$ correspond to the incoming and outgoing electron with mass $m_e$. We assume the polariton dispersion is quadratic, given by $E_k = \hbar^2 k^2 / 2m_p$. The polaritons are taken to be 1-dimensional due to their light mass and the electrons are assume to be 2-dimensional. The 
$\delta$-function accounts for the  energy conservation in the polariton-electron scattering, $N_p(k-k_0)$ is the occupation of the polaritons, and $N_e(\left | \vec{q}-\vec{q}_0 \right |)$ is the occupation of the electrons, where $\vec{q}_0$ is the drift wavevector of the electrons induced by the voltage difference and $k_0$ is the drift wavevector of the polaritons induced by ballistically expanding away from the excitation region. A detailed derivation of this model is discussed in the Supplemental Material~\cite{SI}. 
\par
\begin{figure}
    \centering
    \includegraphics[width=0.45\textwidth]{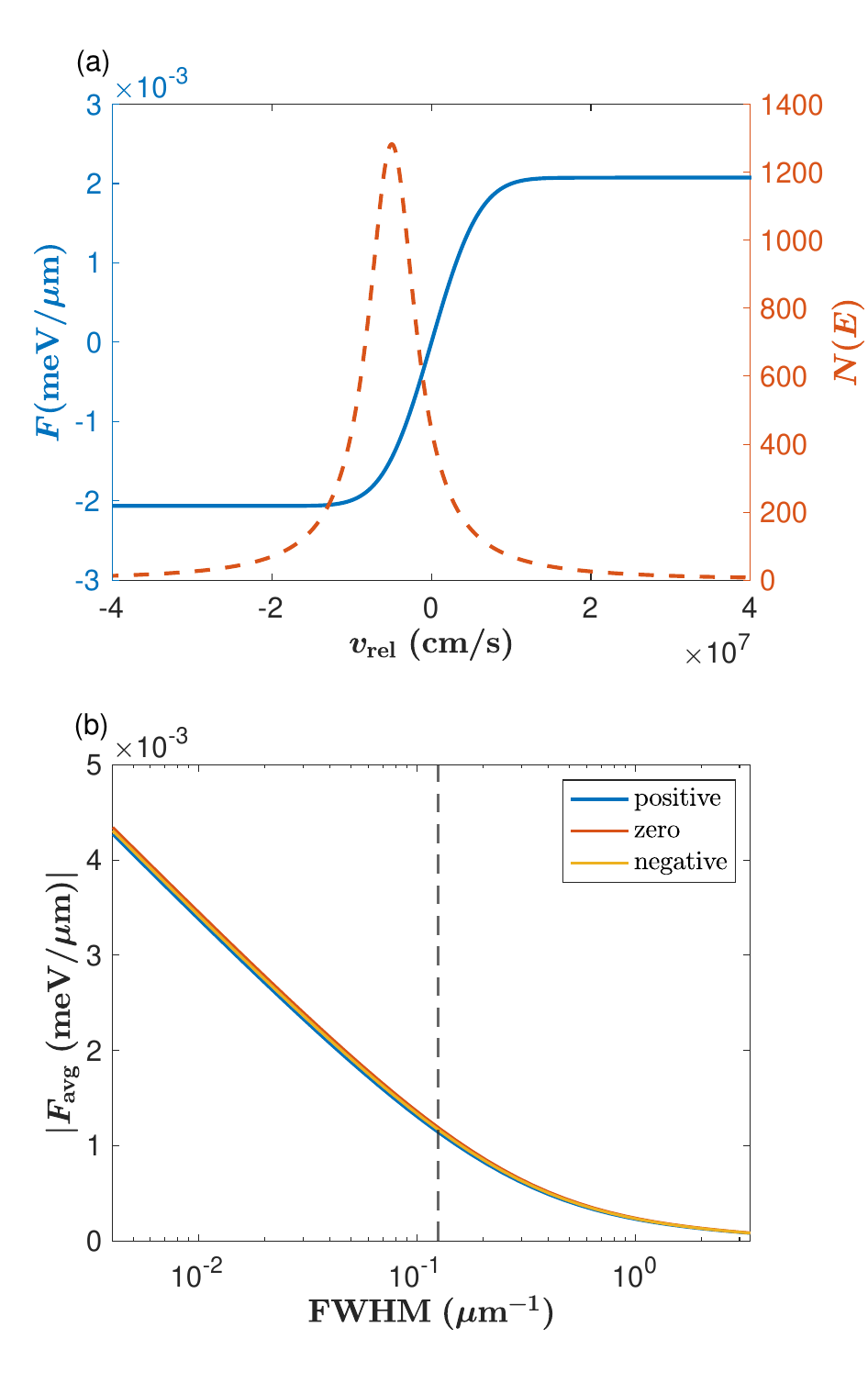}
    \caption{(a) Blue curve: the force the polaritons feel given by Eq.~\eqref{eq:F_k} as a function of the velocity of the electrons relative to the polaritons, defined as $v_{\mathrm{rel}} = \hbar\left ( q_{0}/m_e - k_1/m_p  \right )$. Red dashed line: the distribution $N(v_{\rm rel})$ for the zero-voltage case, extracted from a best fit to the red curve in Fig~\ref{fig:EvK}. (b) The average magnitude of the force the polaritons feel given by Eq.~\eqref{eq:F_avg} as a function of the FWHM of the distribution $N(k)$ of the polaritons, which narrows as their density enters the the quantum degenerate regime. Blue curve: electrons move along the same direction as polaritons. Red curve: electrons are stationary.  Yellow curve: electrons move opposite to polaritons. In all three cases, the polaritons move along the positive $x$-direction but with different speeds, which are extracted from the occupation of the polaritons in Fig.~\ref{fig:EvK}. The vertical dashed line corresponds to the FWHM of the polaritons occupation in Fig.~\ref{fig:EvK}. The parameters used to generate these plots are given in the Supplementary Material.~\cite{SI}}
    \label{fig:drag_thoery}
\end{figure}
Figure \ref{fig:drag_thoery}(a) shows the force polaritons feel due to electron-polariton scattering as a function of the relative velocity between the electrons and polaritons, defined as $v_{\mathrm{rel}} = \hbar\left ( q_{0}/m_e - k_1/m_p  \right )$. When the relative velocity is zero, the polaritons feel no drag force from the electrons. Depending on the sign of the relative velocity, the polaritons can either feel a positive force (push) or a negative force (drag) due to collisions with electrons.
\par

To simulate the experimental data, we fit the three different steady-state $N(k)$ curves in Fig.~\ref{fig:EvK} with a drifted Bose-Einstein distribution $N\left (k-k_0  \right )  = [e^{(E_{k-k_0}-\mu)/k_{B}T} -1]^{-1}$, and extracted the steady state drift momentum $k_0$ of the polaritons for the three different voltage cases (see Supplemental Material~\cite{SI} for more details). We then computed the average force the drifted polaritons feel for the three different voltage cases, which is defined as
\begin{equation}\label{eq:F_avg}
F_{\text{avg}} = \frac{\int_{-\infty}^{\infty} \mathrm{d}k_1\;F(k_1)N(k_1-k_0)  }{\int_{-\infty}^{\infty} \mathrm{d}k_1\; N(k_1-k_0)}.
\end{equation}

As seen in Figure~\ref{fig:drag_thoery}(b), the drag force is a function of the full width at half maximum (FWHM) of the polariton distribution $N(k)$. If the distribution is very broad, then it will feel nearly equal, canceling contributions from electrons moving in opposite directions. However, if the distribution is narrow and offset from zero momentum, as in our experimental $N(k)$, plotted as the dashed red line in Fig.~\ref{fig:drag_thoery}(a), there is a significant net force of the electrons on the polaritons. The theoretical model describes the steady-state velocity regime, which dominates the polariton dynamics. We note that the experimentally extracted velocity corresponds to the measured polariton steady-state velocity. Initially, the polaritons begin with a higher initial velocity, which is then slowed down by drag, through interactions with electrons and other quasiparticles in the system, ultimately reaching the steady-state value we observe.
\par

To fit the data, we assume that the driving force on the polaritons, which is a combination of gradient of the exciton cloud potential and the cavity gradient, is the same in all three cases. This is consistent with our experimental observation that the exciton cloud profile is not affected by the applied voltage, as seen in spatially resolved images presented in the Supplementary material~\cite{SI}. As discussed in the Supplementary material, although there are slight energy differences (ca.~20 $\mu$eV) of the polariton energy in different cases, these appear as changes of the overall average energy and not changes in the gradient; these energy changes are also not correlated with the direction of the current. Using the experimentally measured polariton velocities, we then adjust the electron velocity in the two voltage-driven cases to give the same drag force as the $V=0$ case to cancel out the driving force in steady state, keeping the ratio of the electron velocities the same as the ratio of the measured currents in the two cases. The fact that we can do this self-consistently confirms that our assumptions are valid, in particular that the electron density remains constant, independent of the voltage, and the driving force is unchanged by the voltage. By contrast,  a related, recent experiment on polariton drag \cite{imamoglu} assumed that the main effect was due to a change of the exciton and electron density gradients. From our fits, the absolute value of the electron velocity is estimated as  $9\times10^5$~cm/s in the case of +14~V, which is consistent with expected increase of the measured mobility at room temperature, discussed above.

\par
Our model indicates that in all three different voltage cases in Fig.~\ref{fig:EvK}, the polaritons are slowed down by interaction with the electrons, since the polaritons are faster than the electrons. In the case where the electrons move along the same direction as the polaritons, this drag force is less than when they move in the opposite direction, which gives the shifts of the momentum shown in Fig.~\ref{fig:EvK}.
\par
Because the photoluminescence intensity, which directly indicates the density of the condensate, was not found to be significantly different for different applied voltages, we can rule out the shifts in the condensate energy due to the polariton-polariton and polariton-exciton interactions. What remains is the effect of the the electron drag in our model.


{\bf Conclusions}. We have demonstrated that a direct current can directly alter the momentum of exciton-polaritons; this has the effect of changing the angle of photon emission. Since the polaritons are effectively renormalized photons, created by photon absorption and ending with photon emission, this polariton drag is, in effect, using a direct electric current to change the momentum of photons, in the spirit of the historical Fizeau experiment of 1851 \cite{fizeau}. As polariton structures move ever closer to practical room temperature devices \cite{SnokeKeelingPT2017}, this basic effect may also be possible in those devices. 

We used a simple kinetic theory to explain the effect of drag of a Bose gas interacting with a thermal, fermionic reservoir. We showed that the Bose gas experiences a drag force, which is a function of the relative velocity between the Bose gas and the fermionic reservoir. This indicates that it is simplistic to say that condensates do not experience drag or dissipation. In steady state, when there is interaction with an incoherent reservoir of non-condensate particles, they experience both.

{\bf Funding:}
The work at Pittsburgh was funded by the Army Research Office (W911NF-15-1-0466) and by the National Science Foundation grants DMR-2004570 and DMR-2306977. 
The work of sample fabrication at Princeton was funded in part by the Gordon and Betty Moore Foundation (GBMF-4420) and by the National Science Foundation MRSEC program through the Princeton Center for Complex Materials (DMR-0819860). 

{\bf Author Contributions:}
The experimental work was carried out by D.M.M and Q.Y. under the supervision of D.W.S. D.M.M and Q.Y. contributed equally to this work. The theoretical model was developed by H.A. with the help of S.M. and D.W.S. The sample was grown by L.N.P and K.W., and was designed and characterized by J.B. D.M.M and Q.Y. designed the devices. The device in the main text was fabricated by Q.Y. with the help of B.O. The device in the Supplemental Material was fabricated by B.O. D.M.M, Q.Y., H.A. and D.W.S have done most of the writing. 

\nocite{shouvik,blochRelax,drag1,drag2,drag3,drag4,drag5,drag6,drag7,drag8,russian}

{\bf Data Availability:}
The data that support the findings of this article are openly available~\cite{dataShare}. 

\bibliography{theBib}

\end{document}


\renewcommand{\thefigure}{S\arabic{figure}}
\renewcommand{\theequation}{S\arabic{equation}}
\renewcommand{\thetable}{S\arabic{table}}
\renewcommand{\thesection}{S\arabic{section}}

\title{Pushing Photons with Electrons: Observation of the Polariton Drag Effect - Supplemental Material}

\author{D. M. Myers}
\thanks{Current address: Kulicke \& Soffa Industries, Inc., Fort Washington, PA 19034, USA}
\affiliation{Department of Physics and Astronomy, University of Pittsburgh, Pittsburgh, PA 15260, USA}

\author{Q. Yao}
\thanks{Current address: Joint Quantum Institute, University of Maryland, College Park, Maryland 20742, USA}
\email{qiyao@umd.edu}
\affiliation{Department of Physics and Astronomy, University of Pittsburgh, Pittsburgh, PA 15260, USA}

\author{H. Alnatah}
\affiliation{Department of Physics and Astronomy, University of Pittsburgh, Pittsburgh, PA 15260, USA}

\author{S. Mukherjee}
\thanks{Current address: Joint Quantum Institute, University of Maryland, College Park, Maryland 20742, USA}
\affiliation{Department of Physics and Astronomy, University of Pittsburgh, Pittsburgh, PA 15260, USA}

\author{B. Ozden}
\thanks{Current address: Department of Physics and Engineering, Penn State Abington, Abington, Pennsylvania 19001, USA}
\affiliation{Department of Physics and Astronomy, University of Pittsburgh, Pittsburgh, PA 15260, USA}

\author{J. Beaumariage}
\affiliation{Department of Physics and Astronomy, University of Pittsburgh, Pittsburgh, PA 15260, USA}
\author{L. N. Pfeiffer}
\affiliation{Department of Electrical Engineering, Princeton University, Princeton, NJ 08544, USA}

\author{K. West}
\affiliation{Department of Electrical Engineering, Princeton University, Princeton, NJ 08544, USA}

\author{D. W. Snoke}
\affiliation{Department of Physics and Astronomy, University of Pittsburgh, Pittsburgh, PA 15260, USA}

\date{\today}

                          
\maketitle
\section{Characteristics of the sample used in the main text}

The sample has the same design as used in Ref.~\onlinecite{Steger2015}, where it was reported that the lower polariton lifetime was about 200 ps, and cavity Q was about 320,000. The quality of samples from different batches from the Princeton laboratory of Pfeiffer and coworkers is stable. Ref.~\onlinecite{shouvik} uses another sample, and it indicates the lifetime of polariton is about 200 ps as well. 

We followed the process in Ref.~\onlinecite{Jonny2024} for sample characterization. The sample was characterized by examining different positions before it was etched. At one particular position of the sample, lower polariton energy (LP energy) was obtained directly from the experiment, while upper polariton energy (UP energy) was determined by photoluminescence excitation (PLE) experiment because the cavity is of very high Q factor and the upper polariton signal couldn't be seen directly. With the growth sheet as reference, a model was set up to simulate the reflectivity of the cavity using transfer-matrix method. The model use several fitting parameters to fit the LP and UP dips of the reflectivity spectrum. These parameters are cavity thickness correction $r$, namely the ratio of the actual cavity layer thickness to the designed thickness (due to growth rate variation of the molecular beam epitaxy system), the exciton energy ($E_{ex}$), the exciton linewidth ($\gamma_{ex}$), and the imaginary part of the complex index of GaAs quantum well ($\kappa$), which is a measure of the intrinsic absorption of the material.

\begin{figure}[h]
    \centering
    \includegraphics[width=0.6\linewidth]{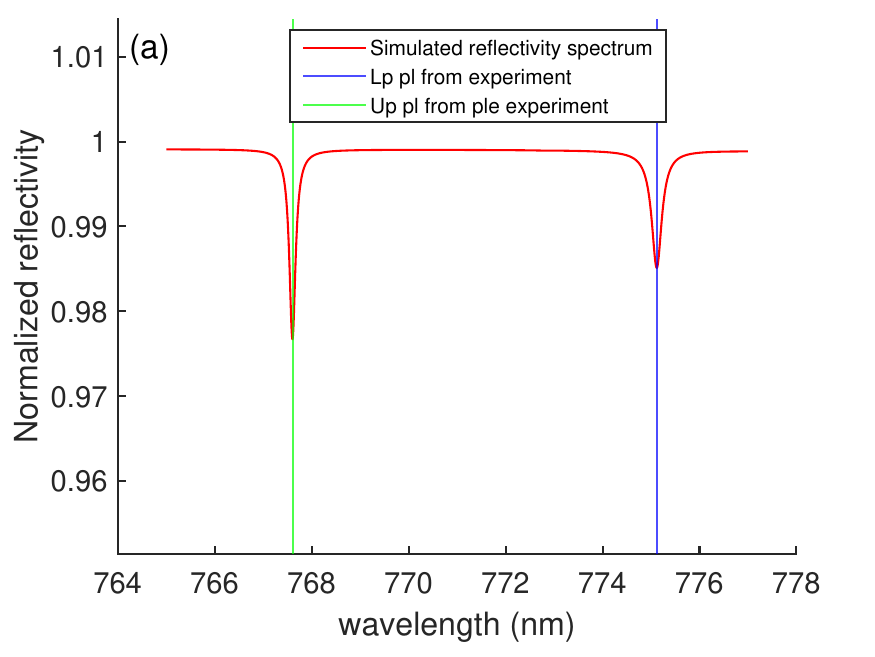}
    \includegraphics[width=0.6\linewidth]{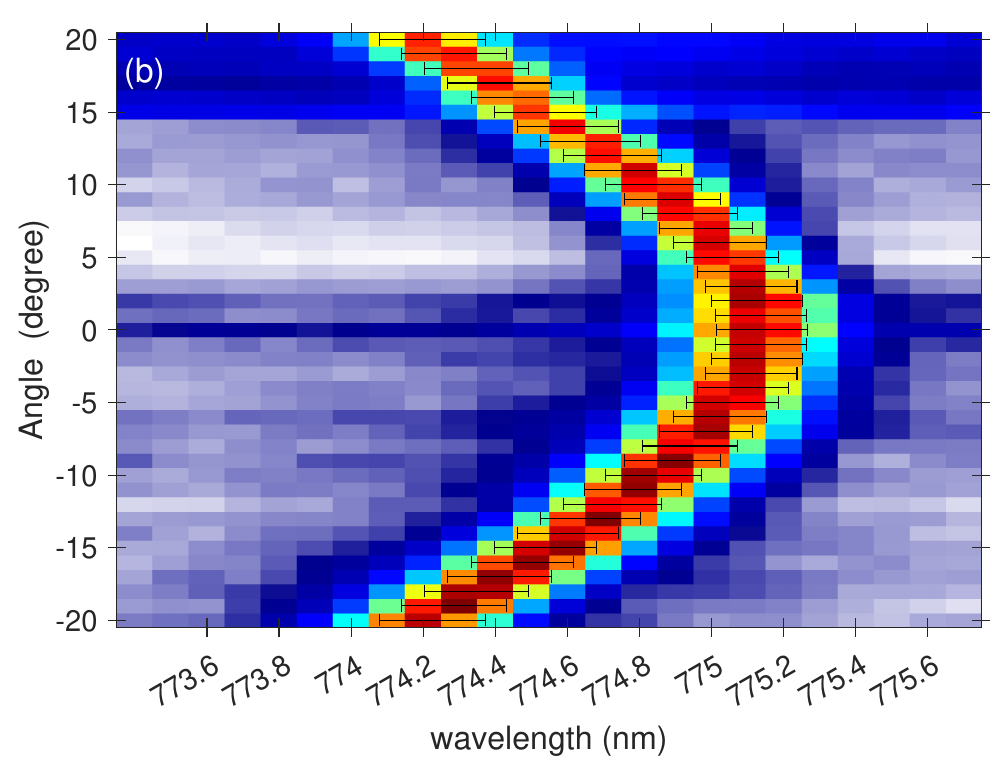}
    \caption{(a) simulated reflectivity curve (red), LP PL (blue) and UP PL (green) at normal incident. (b) Simulation on top of experimental LP PL. The black bars labels the simulated linewidth of LP.}
    \label{fig:sampleSim}
\end{figure}

Figure~\ref{fig:sampleSim} is the characterization result of one position where the LP PL is close to that in the main text. Figure~\ref{fig:sampleSim}(a) shows the simulated reflectivity curve (red) at normal incidence. The dips of the simulated curve aligned very well with experimental PL from the LP (blue) and the UP (green) data. Figure~\ref{fig:sampleSim}(b) shows the simulation on top of the experimental LP PL. The black bars label the simulated linewidth of the LP at different angles. The fitting parameters used in the simulation are: $r = 5.39\%$, $E_{ex} = 1604.40$ meV, $\gamma_{ex} = 0.71$ meV, $\kappa = 0.018$. Based on the those parameters, the exciton fraction of polariton at 1599.7 meV is 66.86\%, and the coupling strength (half the LP-UP splitting) is 8.4 meV. With these parameters, the calculated $Q$ of the bare cavity is 335,000, which is comparable to the previous experiment. 


The gradient of the bare lower polariton energy of the device can be obtained by comparing the LP energy at each end of the device for very low excitation density. In the experiment, lower polariton was generated by low power laser to avoid energy blue shift. The energy difference between two ends of the device used in these experiments was 0.5 meV. Since the wire device is 200 $\mu$m long, the energy gradient is 0.0025 meV/$\mu$m.

\begin{figure}[h]
    \centering
    \includegraphics[width=0.35\linewidth]{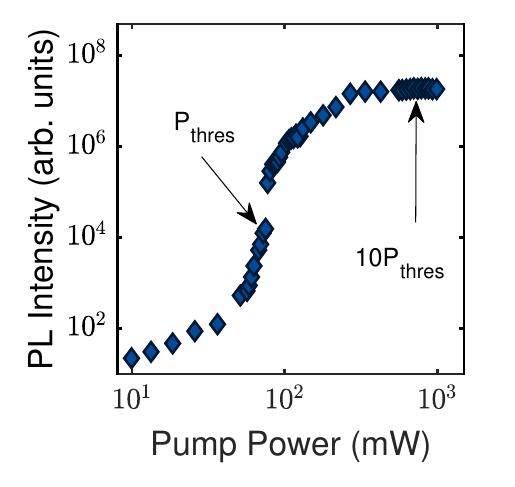}
    \caption{Log-log plot of the photoluminescence (PL) intensity versus the pump power for the device in the main text. Data was taken by pumping on the left side of the devices in reference to Figure 1(b) of the main text. Threshold power, $P_\mathrm{thres}$, and pump power, 10 $P_\mathrm{thres}$ are labeled.}
    \label{fig:Pseries}
\end{figure}

Figure~\ref{fig:Pseries} shows dependence of the photoluminescence (PL) intensity vs.\ the pump power for the device. Threshold power is defined as the power where the non-linearity of the curve begins. As it shows in Figure~\ref{fig:Pseries}, the threshold power was $75$ mW for this device. The data in Figure~\ref{fig:Pseries} was taken when pumping at the left side of the devices in reference to Figure 1(b) of the main text. 

\begin{figure}[h]
\centering
\includegraphics[width=\linewidth]{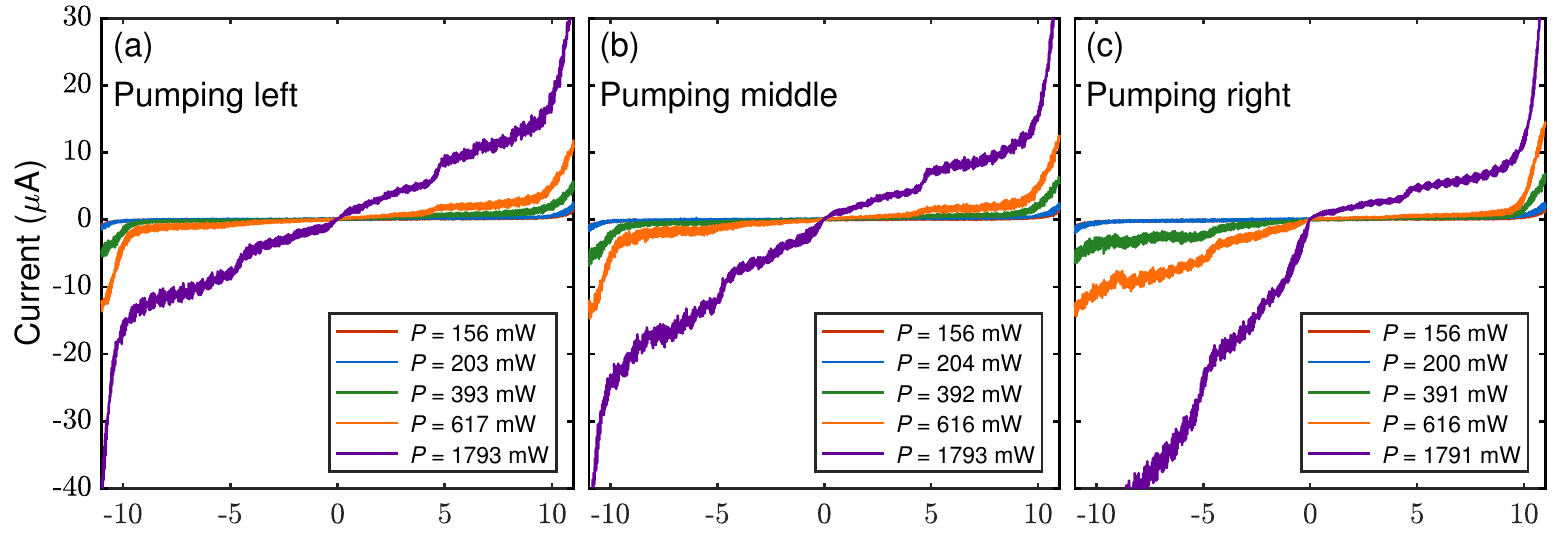}
\includegraphics[width=\linewidth]{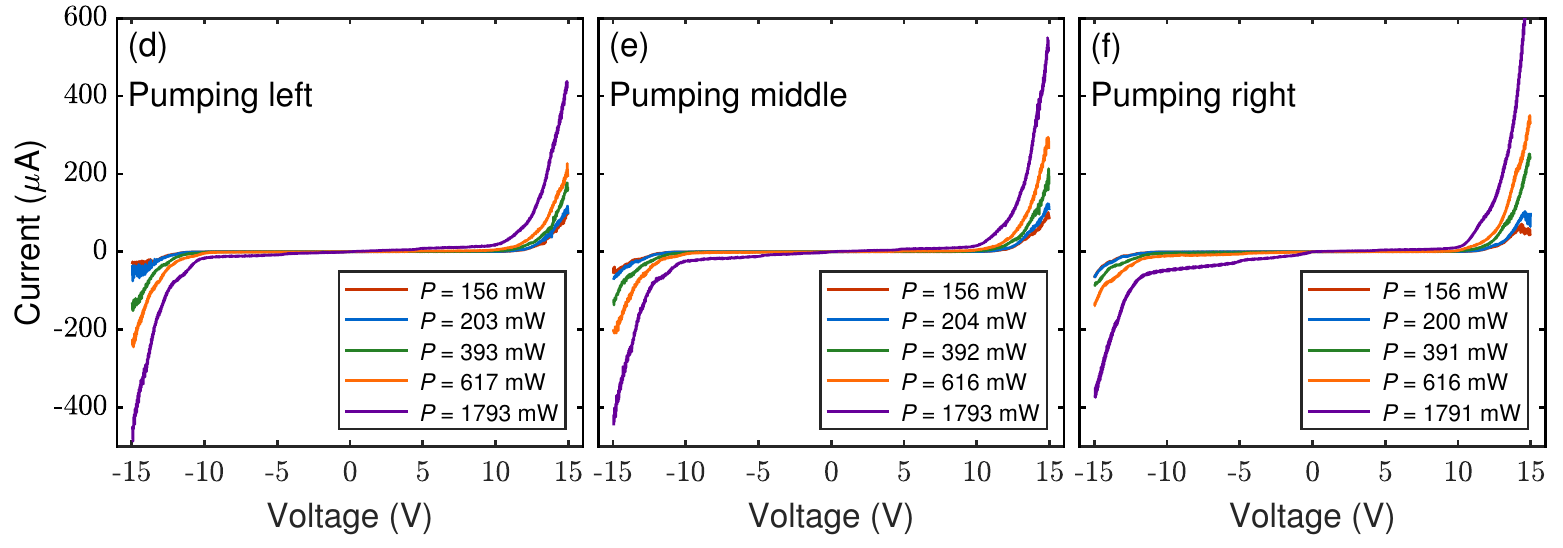}
\caption{Current versus voltage for various pump powers and with different pump spot locations. The locations are given in the upper left of each plot, and are defined in reference to Figure 1(b) of the main text. Specifically, they are the (a)(d) low energy end, (b)(e) middle, and (c)(f) high energy end of the wire, which correspond to the left, middle, and right sides of the $x$-axes used throughout this work. (a), (b), (c) focus on the linear part of (d), (e), (f).}
\label{fig:IV}
\end{figure}

The dark current of the device was $0.8$ nA. The current versus voltage characteristic is similar to what was observed in the square pillar devices of Ref.~\onlinecite{MyersPRB2018}, and is shown in Figure~\ref{fig:IV}. In Figure~\ref{fig:IV}(a), (b), (c), an overall asymmetry is apparent when pumping on either end of the wire, and especially when pumping at the right. 
The asymmetry is nearly opposite for pumping on opposite ends, which indicates that it is primarily due to the pump spot location. We attribute this to greater illumination of the contact near the pump spot, creating free carriers that can carry current over the $n-i$ band bending barrier, as discussed in Ref.~\onlinecite{MyersPRB2018}. Figure~\ref{fig:IV}(d), (e), (f) include wider voltage range. Significant nonlinear rise of the current can be observed when applying voltages value above 10 V.

\newpage

\section{Additional data showing the drag effect}

\begin{figure}[h]
    \centering
    \includegraphics[width=\linewidth]{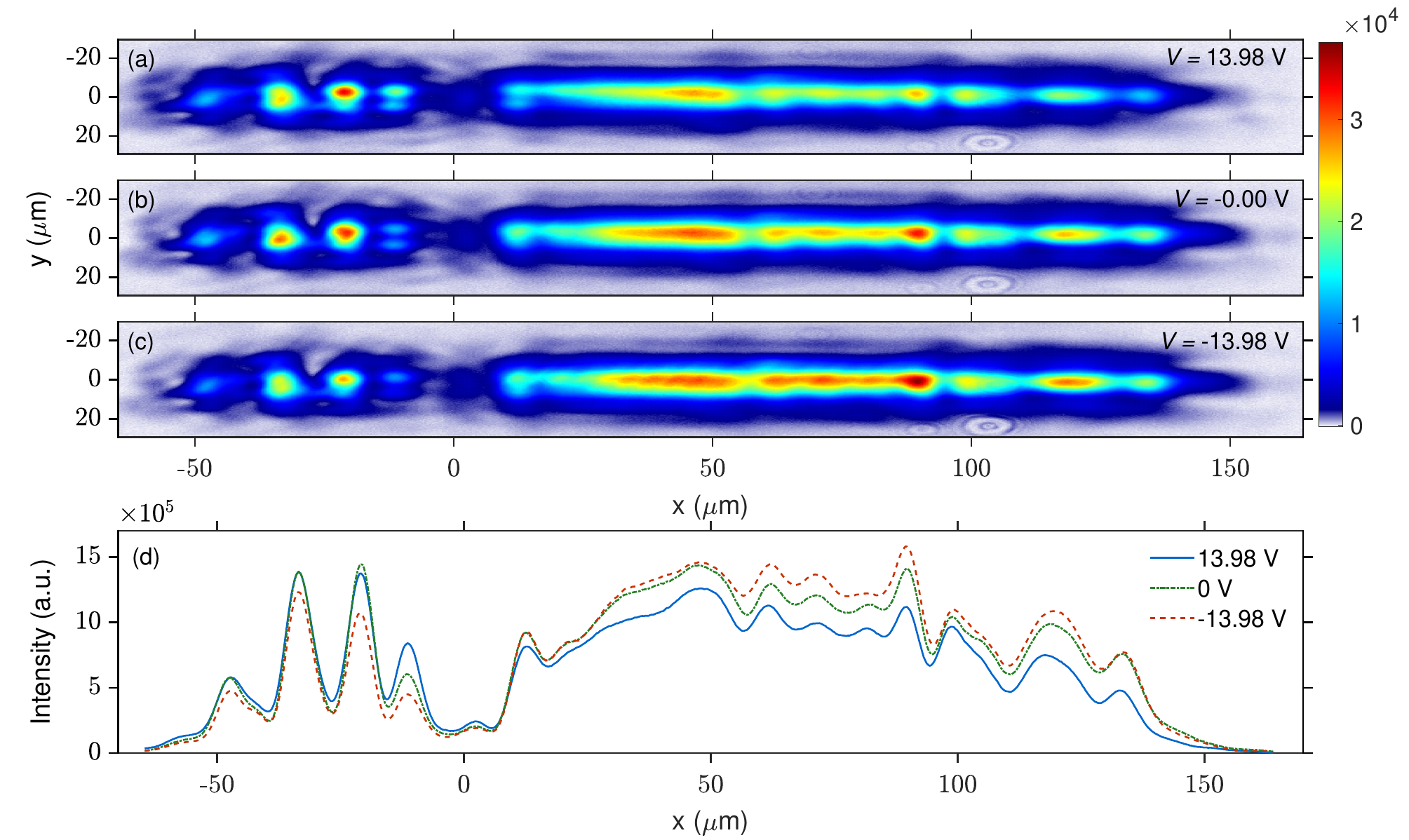}
    \caption{Spacial distribution of the polariton condensate in the wire, under three different applied voltages. The threshold power ($P_\mathrm{thres}$) was about 75 mW, and the powers used were about 10.3$P_\mathrm{thres}$. The pump spot was at $x=0\; \mathrm{\mu m}$, but its image was filtered out by a 750  nm longpass filter. The applied voltage in each case (a) 14~V and (b) 0~V (c) -14~V. (d) Integrated intensity vs.\ position along the wire extracted from (a)-(c).}
    \label{fig:R}
\end{figure}

Fig.~\ref{fig:R} shows the real-space images of the condensate under different applied voltages, with the same experimental condition as Fig.~3(a)-(c) in the main text. Although the current changed the momentum of the polariton condensate, it didn't change its spatial distribution significantly. Figure~\ref{fig:R}(d) shows that the main effect of the current is to change the overall population of the polaritons. The change in population density is consistent with the assumption that there is a region of faster decay near the ends of the wire, presumably due to the defects introduced by the contacts there. When the electrons give an opposing force to the polariton motion toward an end of the wire, that keeps them away from the region of faster recombination at the end of the wire. This gives an increase in overall lifetime that increases the population by about 10\%. Conversely, if the electrons speed up the travel toward the end of the wire, it decreases the population by about 10\%. This effect is seen at both ends of the wire, for polaritons traveling in opposite directions away from the creation point at $x=0$.

\begin{figure}[htb]
\centering
\includegraphics[width=\linewidth]{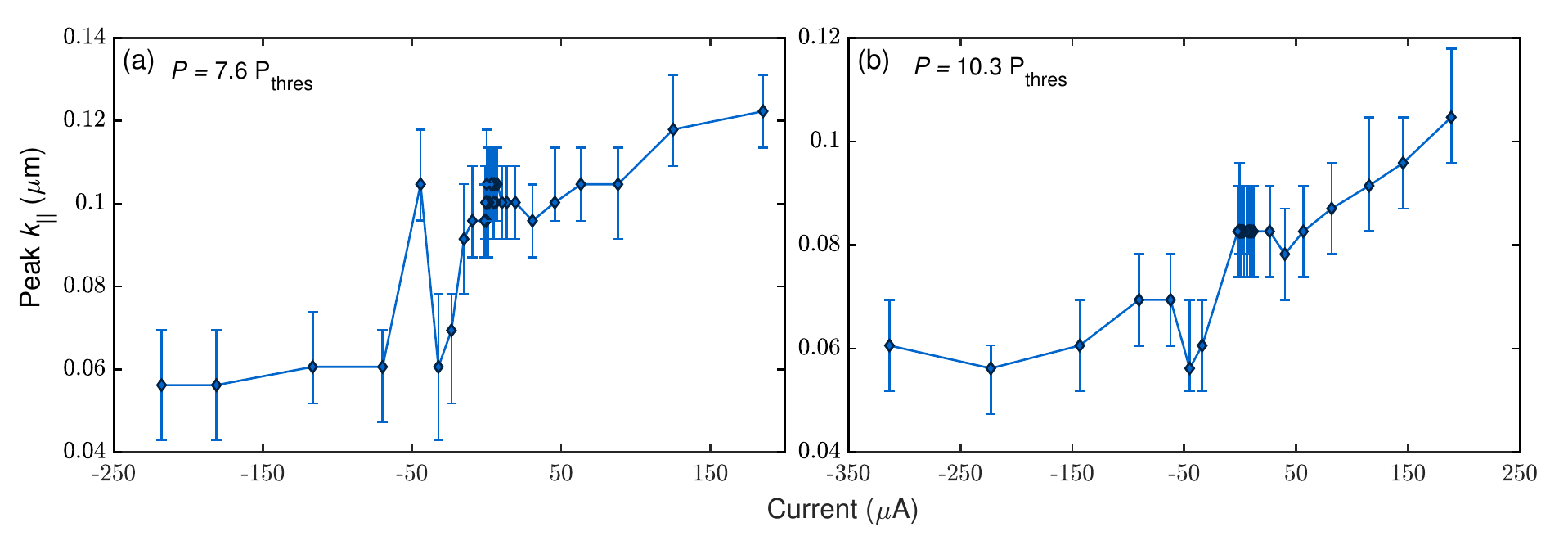}
\caption{Peak value of $k_{||}$ as a function of current, extracted from images similar to Fig.~3(b) for the device used in the main text. The threshold power for this device was 75 mW, as mentioned in the previous section. Error bar labels $k_{||}$ at 95\% of the peak. (a) laser excitation power of $7.6$ $P_{thres}$. (b) laser excitation power of $10.3$ $P_{thres}$.}
\label{fig:KvI}
\end{figure}

Figure~\ref{fig:KvI} shows the peak value of $k_{||}$ versus current for the device used in the main text, extracted from images similar to Fig.~3(b) in the main text. Figure~\ref{fig:KvI}(a) was taken with laser excitation power of $7.6$ $P_{thres}$, and Figure~\ref{fig:KvI}(b) was taken with laser excitation power of $10.3$ $P_{thres}$. Error bar labeled $k_{||}$ position at 95\% of the peak.  

\begin{figure}[htb]
\centering
\includegraphics[width=0.8\linewidth]{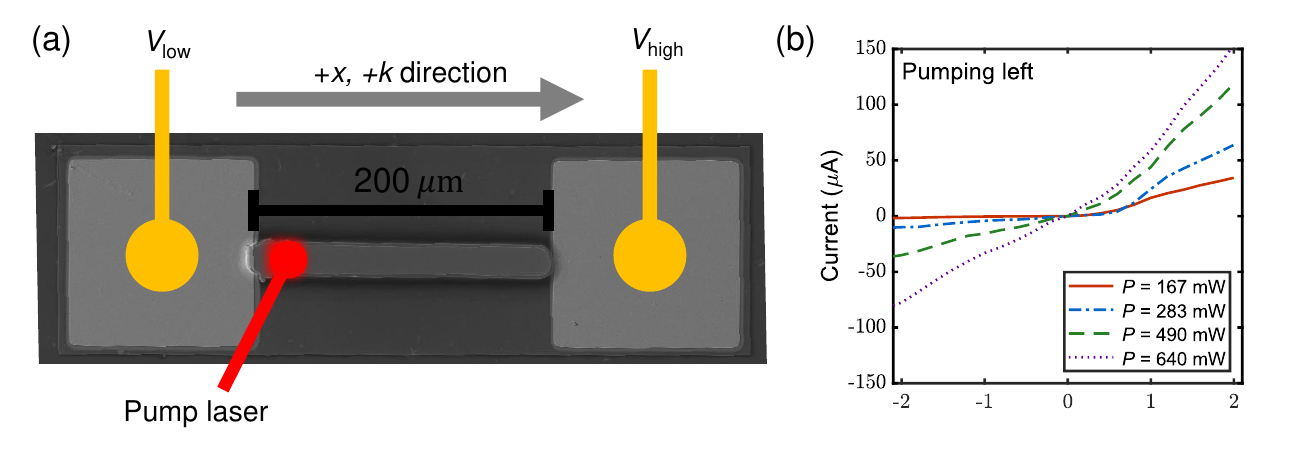}
\includegraphics[width=\linewidth]{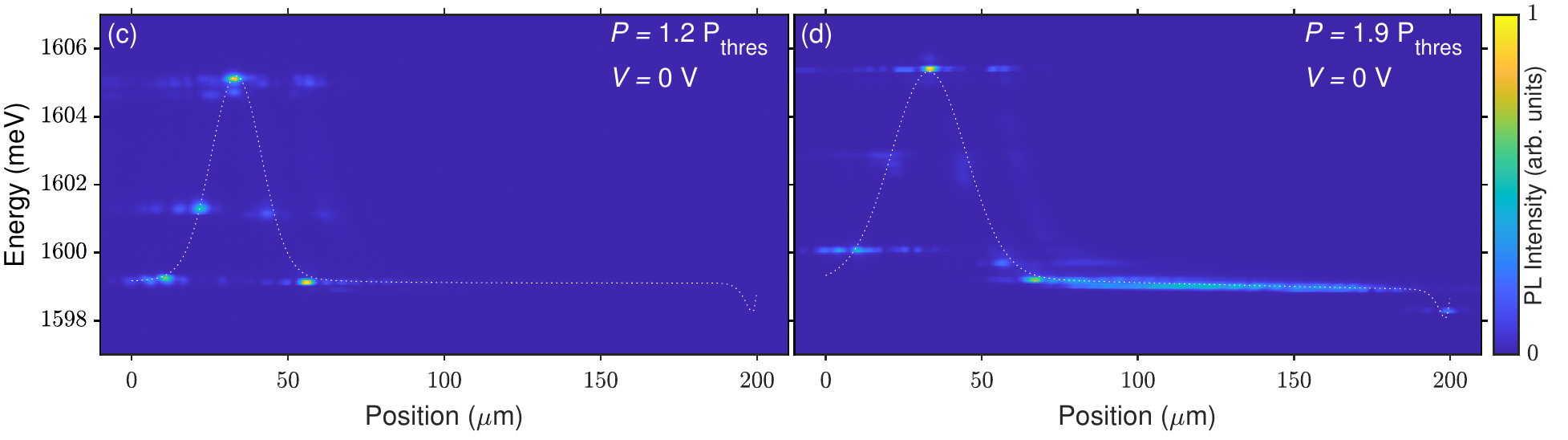}
\caption{(a) Scanning electron microscope (SEM) image of a more photonic device, with an overlay showing the same experimental arrangement as Figure 1 in the maintext. (b) Current-voltage characteristic of this device when pumping on the left side of the wire. (c), (d) PL intensity versus energy and position along the length of the wire at zero applied voltage. The white dotted lines give the outline of the potential felt by the polaritons. The threshold power ($P_\mathrm{thres}$) was about 130 mW, and the powers used were (c) 1.2 $P_\mathrm{thres}$ and (d) 1.9 $P_\mathrm{thres}$. The PL intensity was normalized separately for each image.}
\label{fig:diagram_SI}
\end{figure}

The drag effect was also observed when pumping at the low energy end of a more photonic device, with detuning -3 meV on the left end and -2.1 meV on the right end. This device has a larger lower polariton energy gradient about 0.0035 meV/$\mu$m. Figure~\ref{fig:diagram_SI}(a) shows the SEM image of the device, with the experimental arrangement same as Figure 1 in the main text. When pumping at the left end of the wire, the threshold power, $P_\mathrm{thres}$, is 130 mW. 
The current-voltage characteristic is presented in Figure~\ref{fig:diagram_SI}(b). Although this device has no transition area at the end of the wire, like the one in the main text, it has much lower resistance (compare Figure~\ref{fig:diagram_SI}(b) to Figure~\ref{fig:IV}), because the metal contact and quantum well formed better ohmic contacts in this device. This allowed us to observe the drag effect on this device at lower pump power and lower voltage compared to the conditions in the main text. Figure~\ref{fig:diagram_SI}(c) and (d) show energy-resolved real-space images at zero applied voltage, similar to Figure 2 in the main text. The dip at the end of the wire is due to the increased strain from the etching, similar to the corners of the square pillars discussed in Ref.~\onlinecite{Myers2017}.

\begin{figure}[t]
\centering
\includegraphics[width=0.48\linewidth]{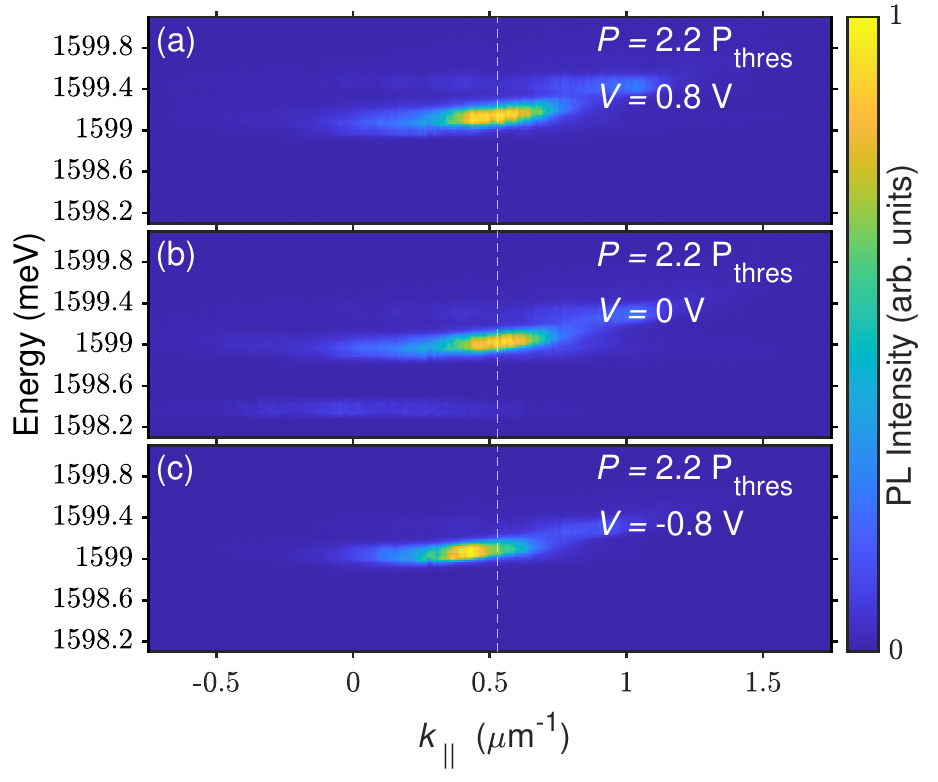}
\includegraphics[width=0.48\linewidth]{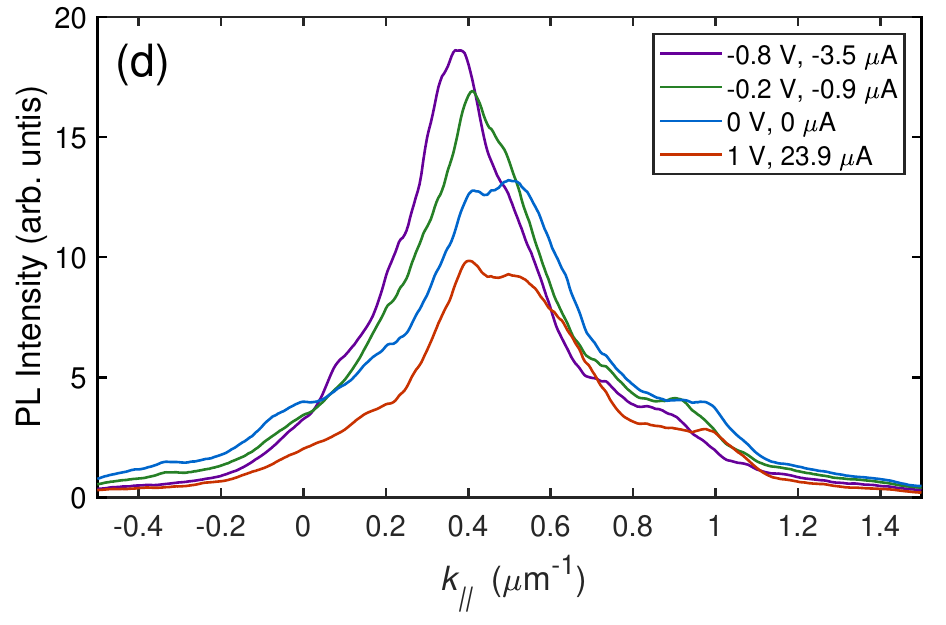}
\caption{(a), (b), (c) PL intensity versus energy and in-plane momentum ($k_\parallel$) for the second sample discussed in this supplementary file. The threshold power ($P_\mathrm{thres}$) was about 130 mW, and the powers used were both about 2.2$P_\mathrm{thres}$. The applied voltage in each case (a) 0.8~V, (b) 0~V and (c) -0.8~V. The PL intensity was normalized separately for each image. (d) Time integrated average momentum distribution of the condensate under different applied voltages derived from data like (a), (b), and (c).}
\label{fig:EvK}
\end{figure}

Figure~\ref{fig:EvK}(a), (b), (c) show the energy versus in-plane momentum ($k_\parallel$) of the polaritons at three different applied voltages, similar to Figure 3 in the main text. The data was collected with the pump region blocked by a real space filter. A population near $k_\parallel = 0$ and at slightly lower energy is visible, and is emitted from the end trap. At zero voltage, the condensate has positive momentum due to flowing towards the $+x$ direction. At negative voltage, the overall momentum is clearly reduced. Figure~\ref{fig:EvK}(d) shows the polariton distribution versus $k_\parallel$ for multiple applied voltages. A clear shift in the distribution toward lower $k_\parallel$ is visible with applied negative voltage, indicating drag upon the polaritons. However, the increase of the condensate momentum was not observed in this device. This is because the polariton velocity in this device is much higher than the device discussed in the main text. In this deviced, the polaritons move approximately 6 times faster in this device than the device shown in the main text. In addition, the electrons move much slower in this device. Using the same calculations as in the main text, the mobility 100~cm$^2$/V-s and voltage drop of 1~V over 100~$\mu$m, the electron velocity is approximately $5\times10^3$~cm/s, an order of magnitude slower than the electrons for the device in the main text. Our theoretical model (discussed below) indicates that the drag force becomes constant when the relative velocity is high $ > 3\times 10^7$ cm/s (see orange curve in Fig. 4 of the main text).

\newpage
\section{Influence of the injected electrons in the pump region and energy gradient}

In our experiment, a non-resonant pump was applied on one side of the wire, making an exciton reservoir. The excitons fell into the lower polariton states due to scattering with the lattice via phonon emission, and relaxed to the lower polariton states near $k_{\mathrm{\parallel}} = 0$ as a result of the polariton-polariton interaction \cite{Microcavities2017}. The polariton-electron interaction can also enhance polariton relaxation process and increase the population near $k_{\mathrm{\parallel}} = 0$ on the lower polariton branch at low polariton density, as shown theoretically \cite{Hartwell} and by direct measurement in experiments \cite{blochRelax}.

\begin{figure}[htb]
\centering
\includegraphics[width=0.48\linewidth]{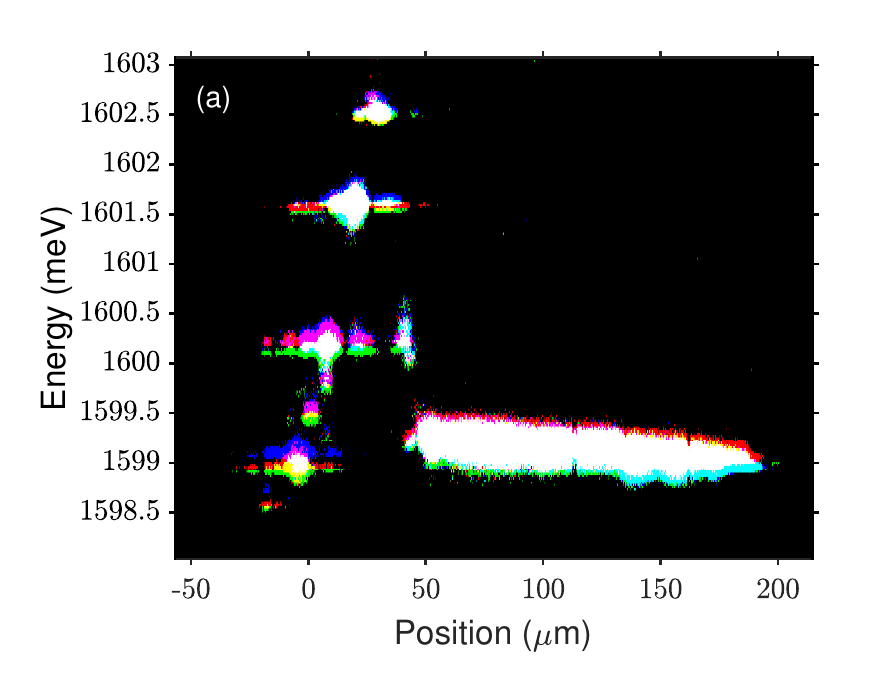}
\includegraphics[width=0.48\linewidth]{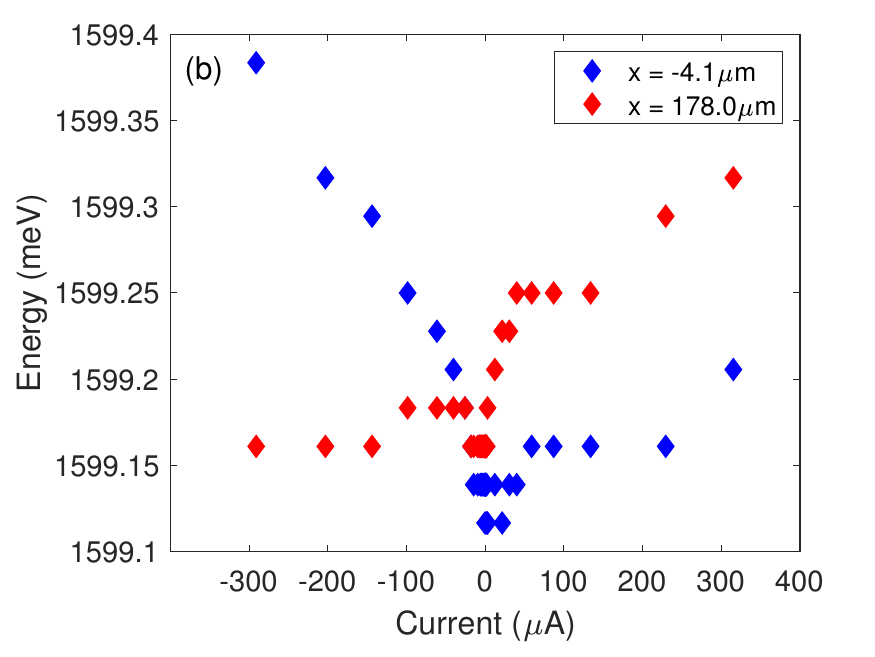}
\includegraphics[width=0.45\linewidth]{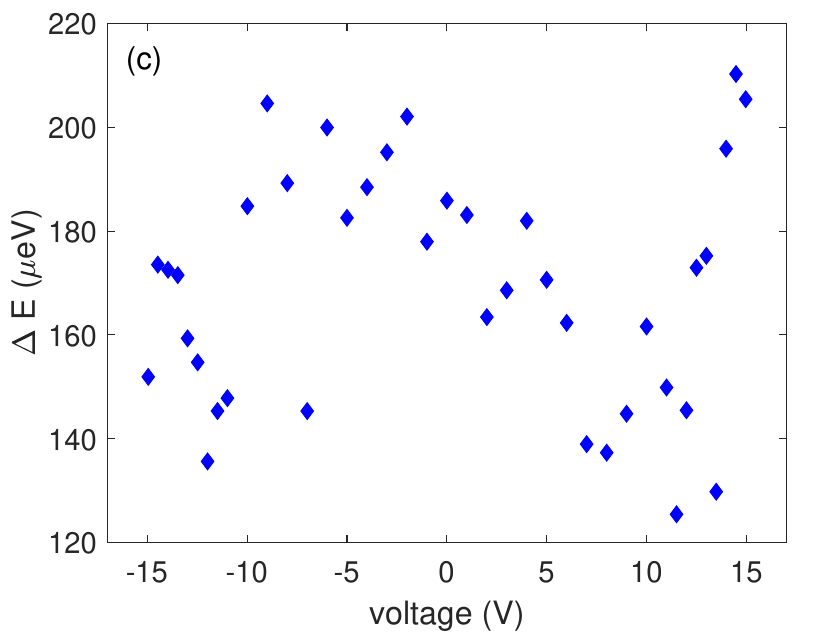}
\caption[pump]{Additional experimental data for the device in the main text. (a) RGB data from energy-resolved real space images with the pump region. The pump power was 10.1$P_\mathrm{thres}$. Red, green, and blue curves correspond to data with applied voltage -14 V, 0 V, and 14 V. Yellow, cyan, and magenta represent overlap of the -14 V and 0 V curves, 0 V and 14 V curves, and -14 V and 14 V curves, respectively. White corresponds to all three overlapping. (b) Energy vs.\ Current at two positions: left-end trap ($x = -4.1$$\mu$m), and near the right end of the wire ($x = 178$$\mu$m). (c) Energy difference between $x = 60.5~\mu$m and $x = 177.7\mu$m, i.e., in the range of the wire in which the change of momentum was measured.}
\label{fig:pumpSpotPaper}
\end{figure}

One might imagine a mechanism by which the enhanced cooling of the excitons and polaritons changed the potential-energy profile at the injection spot, which could change the speed of the escaping polaritons. From general considerations, this is unlikely, because the cooling rate is a function of the density of the carriers, not the speed of the current, to first order. However, we can look into this possibility in more detail by examining the spatial profile of the pump spot as voltage was changed. 

Figure~\ref{fig:pumpSpotPaper}(a) shows data for the energy versus spatial position with three applied voltages for the device in the main text. False color has been used with red, green, and blue corresponding to data with applied voltages -14 V, 0 V and 14 V, respectively; regions of overlap are indicated by other colors as given in the caption. As seen in this figure, the energy position of the condensate at the top of the profile (1602.5 meV) is not observably changed by the applied voltage, indicating that the injection velocity from the top of the excitation profile will be unaffected by voltage.

Another observation is that the polariton energy changed due to the interaction between exciton reservoir and the injected electrons. Figure~\ref{fig:pumpSpotPaper}(b) gives energy change due to current on the left ($x = -4.1\mu$m), and near the end of the wire on the right ($x = 178\mu$m). However, the energy gradient of the region of interest ($x = 60 - 180\mu$m) was almost constant. Figure~\ref{fig:pumpSpotPaper}(c) shows the energy difference between two positions with various applied voltages. Also, the energy change was not observed in the more photonic device, which was discussed in the previous section.

\section{Theory of Polariton Drag}
We consider the polariton-electron scattering to derive the drag force that the electrons exert on the polaritons. The out-scattering rate for a polariton with momentum $k_1$ follow Fermi’s golden rule
\begin{equation}
\frac{\partial N_{\vec{k}_1}}{\partial t} = \frac{2\pi}{\hbar}\sum_{k_f}\left | \left \langle \vec{k}_f|V_{\text{int}}|\vec{k}_i \right \rangle \right |^2\delta\left ( E_{\text{f}}-E_{\text{i}} \right ).
\end{equation}
Here, we consider the polariton-electron scattering where the two particles scatter from the state $\ket{\vec{k}_i} = \ket{k_1\vec{q}_1}$ to state $\ket{\vec{k}_f} = \ket{k_2\vec{q}_2}$ through a scattering interaction $V_{\text{int}}$. The wavevectors $k_1$ and $k_2$ correspond to the incoming and outgoing polariton with mass $m_p$, respectively and $\vec{q}_1$ and $\vec{q}_2$ correspond to the incoming and outgoing electron with mass $m_e$. We take the electrons to be 2-dimensional but the polaritons are treated to be 1-dimensional. This is because the confinement energy of the wire is inversely proportional to the mass, making polaritons feel much more confinement than the electrons since the polaritons are much lighter. The interaction potential for the polariton-electron scattering is given by
\begin{equation}\label{eq:Vint}
V_{\text{int}} = V_{0}\sum_{k_1,k_2,\vec{q}_1}a^{\dagger}_{k_2}b^{\dagger}_{\vec{q}_2}b_{\vec{q}_1}a_{k_1},    
\end{equation}
where $a^{\dagger}$ and $a$ are the polariton creation and annihilation operators respectively and $b^{\dagger}$ and $b$ are the electron creation and annihilation. Substituting Eq.~\eqref{eq:Vint} into Fermi’s Golden rule, we obtain
\begin{equation}
\frac{\partial N_p({k_1})}{\partial t}=\frac{2 \pi}{\hbar}V^2_{0} \sum_{\vec{k}_2, \vec{q}_2} N_e( |  \vec{q}_1 |)\left ( 1-N_e( |  \vec{q}_2 |) \right )\;N_p(k_1)\left ( 1+N_p(k_2) \right )  \delta\left(E_{k_2}+E^e_{\vec{q}_2}-E_{k_1}-E^e_{\vec{q}_1}\right),
\end{equation}
where $N_e( |  \vec{q} |)$ is the occupation number of the electrons and $N_p(k)$ is the occupation number of the polaritons. Converting the sum to an integral gives
\begin{equation}\label{rate_k1}
\begin{split}
\frac{\partial N_p({k_1})}{\partial t}=\frac{2 \pi}{\hbar}V^2_{0} \frac{A}{(2\pi)^2}\frac{L}{2\pi}\int \int\mathrm{d}q_{2_x}\;\mathrm{d}q_{2_y} \int\mathrm{d}k_{2}\; N_e( |  \vec{q}_1 |)\left ( 1-N_e( |  \vec{q}_2 |) \right )N_p(k_1)\left ( 1+N_p(k_2) \right )\\
\times \delta\left(E_{k_2}+E^e_{\vec{q}_2}-E_{k_1}-E^e_{\vec{q}_1}\right), 
\end{split}
\end{equation}
where we have taken the polaritons to be moving along the $x$-direction. Here, $L$ is the length of the wire and $A = L\times W$ is the area of the wire, $q_{2_x}$ and $q_{2_y}$ are the $x$ and $y$ components of the outgoing electron wavevector and $k_{2}$ is the outgoing 1-dimensional polariton wavevector, which is taken to be along the $x$-direction. Rewriting Eq.~\eqref{rate_k1} gives
\begin{equation}
\begin{split}
\Gamma({k_1}) = \frac{1}{N_p(k_1)}\frac{\partial N_p({k_1})}{\partial t}= \frac{g^2}{W \hbar(2\pi)^2}\int \int\mathrm{d}q_{2_x}\;\mathrm{d}q_{2_y} \int\mathrm{d}k_{2_x}\; N_e( |  \vec{q}_1 |)\left ( 1-N_e( |  \vec{q}_2 |) \right )\left ( 1+N_p(k_2) \right ) \\
\times \delta\left(E_{k_2}+E^e_{\vec{q}_2}-E_{k_1}-E^e_{\vec{q}_1}\right), 
\end{split}
\end{equation}
where $\Gamma_{k_1}$ is the out-scattering rate and $g^2 = V^2_0A^2$. Since the goal is to compute the shift of polariton momentum, we weigh each scattering by the change of momentum of the polariton. 
\begin{equation}
\begin{split}
\hbar \Delta k \Gamma({k_1}) = \frac{g^2}{W \hbar(2\pi)^2}\int \int\mathrm{d}q_{2_x}\;\mathrm{d}q_{2_y} \int\mathrm{d}k_{2}\;\hbar (k_2-k_1) N_e( |  \vec{q}_1 |)\left ( 1-N_e(q_2) \right )\left ( 1+N_p(k_2) \right )\\
\times \delta\left(E_{k_2}+E^e_{\vec{q}_2}-E_{k_1}-E^e_{\vec{q}_1}\right).
\end{split}
\end{equation}
The term $\hbar \Delta k \Gamma({k_1})$ gives the the force the electrons are exerting on a polariton with momentum $k_1$, which we will denote with $F(k_1)$. Taking the low density limit of the electrons implies $N_e( |  \vec{q}_1 |)\left ( 1-N_e( |  \vec{q}_2 |) \right )\simeq N_e( |  \vec{q}_1 |)$, which gives
\begin{equation}
F(k_1) = \frac{g^2}{W \hbar(2\pi)^2}\int \int\mathrm{d}q_{2_x}\;\mathrm{d}q_{2_y} \int\mathrm{d}k_{2}\;\hbar(k_2-k_1) N_e( |  \vec{q}_1 |)\left ( 1+N_p(k_2) \right )  \delta\left(E_{k_2}+E^e_{\vec{q}_2}-E_{k_1}-E^e_{\vec{q}_1}\right),
\end{equation}
The argument of the delta function can be written as
\begin{equation}\label{delta_function}
\Delta E = E_{k_2}+E_{\vec{q}_2}-E_{k_1}-E_{\vec{q}_1} = E(k_2)-E(k_1)+\frac{\hbar^2}{2 m_e}\left(q^2_{2_x}+q^2_{2_y}-q^2_{1_x}-q^2_{1_y}\right).  
\end{equation}Momentum conservation along $x$ and $y$ directions implies that $q_{1_x} = q_{2_x}+k_2 - k_1$ and $q_{1_y} = q_{2_y}$, which gives
\begin{equation}
\Delta E = E(k_2)-E(k_1)+\frac{\hbar^2}{2 m_e}\left(q^2_{2_x}-\left ( q_{2_x}+k_2 - k_1 \right )^2\right), 
\end{equation}
where $E(k) = \hbar^2k^2/2m$ is the dispersion of the polaritons. Integrating over the delta function to eliminate $q_{2_x}$ gives
\begin{equation}
 F(k_1) = \frac{g^2m_e}{2\pi^2\hbar^2 W}\int\mathrm{d}q_{2_y} \int\mathrm{d}k_{2}\;\frac{k_2-k_1}{\left | k_2-k_1\right |} N_e\left ( \sqrt{q^2_{1x}+q^2_{2_y}} \right )\left ( 1+N_p(k_2) \right ),  
\end{equation}
where $q_{1_x} = q^{(0)}_{2_x}+k_2 - k_1$. Here, $q^{(0)}_{2_x}$ is given by the roots of Eq.~\eqref{delta_function}, which is given by
\begin{equation}
q^{(0)}_{2_x} = \frac{1}{2}\left ( k_1-k_2 \right )+\frac{m_e}{\hbar^2}\left ( \frac{ E(k_1)-E(k_2)}{k_1-k_2} \right ),    
\end{equation}
and therefore
\begin{equation}
q_{1_x} = \frac{1}{2}\left ( k_2-k_1 \right )+\frac{m_e}{\hbar^2}\left ( \frac{ E(k_1)-E(k_2)}{k_1-k_2} \right ).    
\end{equation}
We assume the electrons are in a drifted Maxwell-Boltzmann distribution, given by  
\begin{equation}
N_e\left ( \sqrt{q^2_{1x}+q^2_{2_y}} \;\right ) = e^{\mu_e/k_{B}T_e}e^{-\hbar^2q^2_{2_y} /2mk_{B}T_e} e^{-\hbar^2\left ( q_{1x}-q_0 \right )^2 /2m_ek_{B}T_e},   
\end{equation}
where $q_0$ is the momentum of the electrons induced by the voltage across the wire, $\mu_e$ is the chemical potential and $T_e$ is the temperature of the electron gas. Integrating over $q_{2_y}$ gives 
\begin{equation}
 F(k_1) = e^{\mu_e/k_{B}T_e}\;\frac{g^2m_e}{2\pi^2\hbar^2 W}\sqrt{\frac{2\pi m_e k_{B}T_e}{\hbar^2}} \int\mathrm{d}k_{2}\;\frac{k_2-k_1}{\left | k_2-k_1\right |} e^{-\frac{\hbar^2\left ( q_{1x}-q_0 \right )^2}{2m_ek_{B}T} }\left ( 1+N_p(k_2) \right ).  
\end{equation}
To account for the fact that polaritons are ballistically expanding
away from the excitation region, we replace $N_p(k_2)$ with $N_p(k_2-k_0)$, where $k_0$ is the momentum induced by exciton cloud. We then have 
\begin{equation}\label{eq:F_k}
 F(k_1) = e^{\mu_e/k_{B}T_e}\;\frac{g^2m_e}{2\pi^2\hbar^2 W}\sqrt{\frac{2\pi m_e k_{B}T_e}{\hbar^2}} \int\mathrm{d}k_{2}\;\frac{k_2-k_1}{\left | k_2-k_1\right |} e^{-\frac{\hbar^2\left ( q_{1x}-q_0 \right )^2}{2m_ek_{B}T} }\left ( 1+N_p(k_2-k_0) \right ).  
\end{equation}
The average force the polaritons feel can then be defined as
\begin{equation}\label{eq:F_avg}
F_{\text{avg}} = \frac{\int_{-\infty}^{\infty} \mathrm{d}k_1\;F(k_1)N_p(k_1-k_0)  }{\int_{-\infty}^{\infty} \mathrm{d}k_1\; N_p(k_1-k_0)}.
\end{equation}
\subsection{Simulating the experimental data}
To simulate the experimental data shown in Fig. 3(d), we fit the PL intensity with the Bose-Einstein distribution $N(k-k_0)$ to extract the drift momentum $k_0$ and the chemical potential of the polaritons. We fixed the temperature of the polaritons to $T = 5\;\mathrm{K}$ and we vary $\mu_p$ and $k_0$ to find the best fit to the Bose-Einstein distribution, given by
\begin{equation}\label{eq:drifted_BE}
N_p(k-k_0) = \frac{A}{e^{\left (E(k-k_0)-\mu_p \right )/k_BT}-1},  
\end{equation}
where $A$, $\mu_p$ and $k_0$ are fit parameters and $E(k-k_0) = \hbar(k-k_0)^2/2m_p$. Since our model does not take into account reflection from the edge of the wire, we exclude the negative momentum PL in Fig. 3(d) from the fit, which comes from reflection of the polaritons off the far end of the wire.
\par
\begin{figure}
    \includegraphics[width=0.6\linewidth]{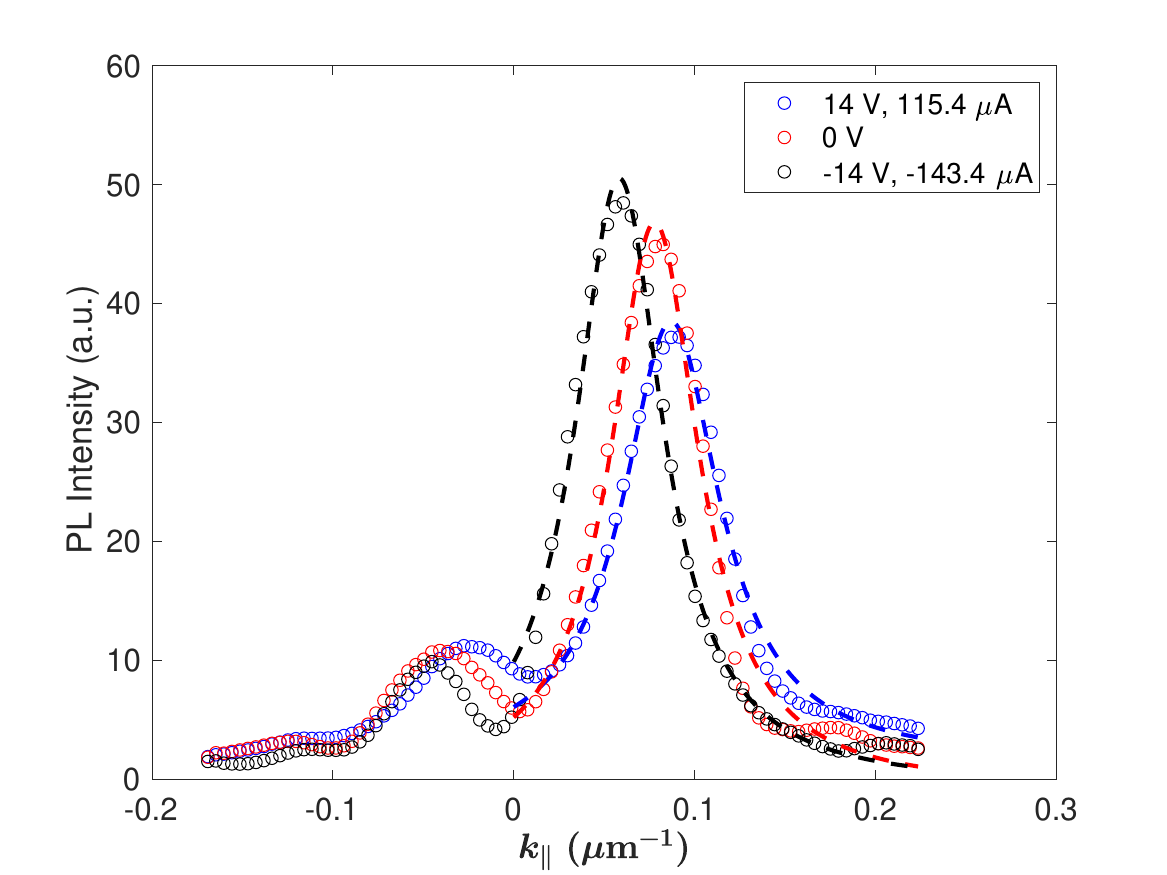}
    \centering
    \caption{Circles: experimental time integrated average momentum distribution of the condensate under different applied voltages (Fig. 3(d) of main text). Dashed lines: the best fit to the drifted Bose-Einstein distribution given by Eq.~\eqref{eq:drifted_BE}. }
    \label{fig:N_k_fit}
\end{figure}

Figure \ref{fig:N_k_fit} shows the fit to the Bose-Einstein distribution for the positive, zero and negative voltages shown in Fig. 3(d) of the main text. The extracted drift momenta $k_0$ from the best fit are $0.0871\; \mathrm{\mu m}^{-1}$, $0.0779\; \mathrm{\mu m}^{-1}$ and $0.0583\; \mathrm{\mu m}^{-1}$ for the $+14$ V, $0$ V and $-14$ V cases respectively. These values are used in the numerics to find the average drag force given by Eq.~\eqref{eq:F_avg} for these three different voltage cases. Since the magnitude of current for the $+14$ V and $-14$ V is not the same experimentally, we assume that the electrons move $143.4\mathrm{\mu A}/115.4\mathrm{\mu A}\approx 1.24$ times faster for the negative voltage case.  

\subsection{Estimation of the magnitude of the drag force}
In this section, we estimate the magnitude of the force that the polaritons feel for a polariton momentum of $7.76\times 10^{4}\; \mathrm{\mu m}^{-1}$ corresponding to the zero voltage case in Fig~\ref{fig:N_k_fit} (i.e. stationary electrons). The electron-polariton interaction strength polaritons has been calculated in Ref.~\onlinecite{li2021theory}. In the low energy and small momentum limit, the polairton-electron interaction is given by $g= 4\pi E_b a^2_b|X|^2$, where $E_b$ is the binding energy of the excitons, $a_b$ is the Bohr radius and $|X|^2$ is the exciton fraction. Here, we use $E_b = 10\; \mathrm{meV}$, $a_b = 13\; \mathrm{nm}$ and $|X|^2 = 0.5$. From Hall measurements, the electron density is $3.89\times 10^9\;\mathrm{cm^{-2}}$. The fits in Fig.~\ref{fig:N_k_fit} for the zero voltage case gives a chemical potential of $\mu_p/k_B T =-9.5\times 10^{-4}$. For these parameters, Eq.~\eqref{eq:F_avg} gives an average force of $F_{avg} = 1.2\times 10^{-3}\; \mathrm{meV/\mu m}$. In steady state, the drag force should balance the cavity gradient force, which we measure experimentally to be $(2.7 \pm 2) \times10^{-3}\; \mathrm{meV/\mu m}$. Despite the many uncertainties in our theoretical estimate, we find that the estimated strength of the force is within the uncertainty of the experimental cavity gradient force. 
 
\section{Comparison to other work}
A series of papers \cite{drag1,drag2,drag3,drag4,drag5,drag6,drag7,drag8} reported observation of a ``photon drag'' effect, also called ``optical rectification'' or a ``photogalvanic effect.'' In these works, a high-intensity laser beam was directed at a semiconductor system, at an optical resonance, and a voltage difference, i.e., an electric polarization, was measured in the material. This can be interpreted as, and was analyzed as, a nonlinear optical effect which is essentially the inverse of the electro-optic effect. In the electro-optic effect, an applied DC electric field leads to a change of the index of refraction of a medium felt by an electromagnetic wave; in these ``optical rectification'' experiments, a nonlinear $\chi^{(3)}$ term led to shift in the index of refraction that then led to a DC electric field. The effect seen in these papers relied on high-intensity lasers to obtain nonlinear effects, while in our experiments, the condensate is a very low photon density. They also relied the presence of a strong absorption resonance, and saw only a voltage in response to a total intensity, not any shift of the momentum of the photons in response to current.

A theoretical paper, Ref.~\onlinecite{russian}, proposed a mechanism by which a current could lead to a drag effect on photons. In that work, it was proposed that a current could shift the subband energies of electron states in a quantum well, which would then shift the index of refraction, which then would have an effect on photons passing through the medium. For that effect to occur, there would have to be substantial shift of the exciton subband states. In our experiments, as seen in Figure~\ref{fig:pumpSpotPaper}, at the point of maximum exciton density, there is no shift of the exciton energy with applied voltage. 

In Ref.~\onlinecite{imamoglu}, an electron density gradient was used to accelerate polaron-polaritons, by creating a voltage-dependent potential energy gradient for the polaritons. Again, as discussed above, the fact that in our experiments the polariton energy at the point of maximum exciton density does not shift at all with voltage, as seen in Figure~\ref{fig:pumpSpotPaper}(a), shows that the free electrons are having negligible effect on the reservoir exciton spatial distribution. There is therefore no effect of the electron current on the static potential energy felt by the polaritons. 

There is a sense in which all of the above experiments and ours may be considered different limits of a general effect, which is that photons and electrons can exchange momentum in a medium with electron-photon resonances. In the experiments reviewed in this section, however, there is intrinsically a much longer time lag, of the order of nanoseconds, as there must be a macroscopic rearrangement of the electron density in the medium to create an electric polarization. In the theoretical model we have presented here, which perfectly describes our experimental results, the response is instantaneous as a result of direct polariton-electron collisions. 

\bibliography{theBib}









